\documentclass[12pt]{article}
\let\section=\subsection     \let\subsection=\subsubsection                
\usepackage{graphicx}
\usepackage{epsfig}

\begin{document}
\begin{center}
   {\large \bf THREE QUARK CLUSTERS IN }\\[2mm]
   {\large \bf HOT AND DENSE NUCLEAR MATTER
}\\[5mm]
   M. Beyer$^a$, S. Mattiello$^a$, T. Frederico$^b$, H.J. Weber$^c$  \\[5mm]
   {\small \it  $^a$Fachbereich Physik, Universit\"at Rostock, 18051
  Rostock, Germany\\
$^b$Dep. de F\'\i sica, ITA, CTA,
12.228-900 S\~ao Jos\'e dos Campos, S\~ao Paulo, Brazil\\
  $^c$Dept. of Physics, University of Virginia, Charlottesville, VA
  22904, U.S.A.}
\end{center}
\begin{abstract}\noindent
  We present a relativistic in-medium three-body equation to study
  correlations in hot and dense quark matter.  The equation is solved
  for a zero-range force for parameters close to the phase transition
  of QCD.
\end{abstract}

\section{Introduction}

Models to study $q\bar q$ bound states, correlations, or condensates
in a hot and dense medium such as nuclear/quark matter are rather
elaborate. These models are used to explore the phase structure of QCD
in particular in areas presently not accessible to lattice
calculations.  Little attention, however, has been paid to $qqq$
correlations in this
context~\cite{Pepin,Beyer:2001bc,Mattiello:2001vq,ropke}. As three
valence quarks are a dominant Fock component in any quark model of
baryons those should be relevant in the confinement-deconfinement
region from nucleons to quarks.
  
Another reason to study three-quark correlations is their relevance
namely $q^3$ correlations to color superconductivity (see e.g.
\cite{Alford:2001}). It has been shown in a different context (Hubbard
model) that the inclusion of three-particle correlations lowers the
critical temperature $T_c$ compared to the bare two-particle Thouless
instability~\cite{letz}.  The existence and the area of the QCD phase
diagram where one might find color superconductivity might depend on
dynamical details and in particular on the influence of three-quark
correlations which have not been investigated so far.

Presently, our aim is to provide a proper framework to tackle the
question of three-body correlations at finite temperatures and
densities for a relativistic many-body system, i.e.  soluble
effective Alt-Grassberger-Sandhas (AGS) or Faddeev equations that
include medium effects as well as special relativity.  To do so,
several problems have to be overcome and in what we present here one
might consider an approach towards a systematic inclusion of medium
effects and relativity in effective three-body equations. In the past
years utilizing the Dyson approach we have derived suitable in-medium
equations to treat three-body correlations as well as four-body bound
states and condensates in a nuclear medium of finite temperatures and
densities, see e.g.~\cite{Beyer:2001iz}. These equations are
generalized here to include special relativity.  To this end we use
the light front approach~\cite{Dirac:49}. Except for details in the
treatment of the medium the use of the light front at finite densities
is similar to that suggested in Ref.~\cite{miller} for $T=0$.

\section{Theory}
For the time being we consider a zero-range interaction. A simple
effective theory of zero range is provided by the Nambu Jona-Lasinio
model~\cite{NJL} that has also been extended to finite temperatures
and densities, see, e.g.~\cite{Klevansky}.  Note that due to screening
effects the effective potential between quarks may loose the confining
property above the critical temperature~\cite{Petreczky:2001pd}.
However, a closer connection to QCD including confinement is certainly
desired, see e.g.~\cite{Pauli:2000np,Frederico:2001qy}. However,
presently, the main focus of our approach is on the structure of the
relativistic in-medium equation.

\subsection{Self energy-correction, gap equation}
In Hartree-Fock approximation the self-energy correction induced by
the medium is given by the gap equation.  The gap equation leads to
effective masses $m(\mu,T)$ that depend on the temperature $T$ and the
chemical potential $\mu$ of the medium. To proceed we use the light
front formalism. The four-dimensional gap equation is projected onto
the light front using the Lepage-Brodsky regularization
scheme~\cite{LB}.
\begin{equation}
m=m_0+2i\lambda N_{\mathrm{c}}N_{\mathrm{f}}\int\limits_{{\rm LB}}
\frac{d^4k}{(2\pi)^4}\;{\rm Tr}\{S(k)\}\;\left(1-{f(k)}\right)
\end{equation}
 The Fermi-Dirac
distribution function
\begin{equation}\label{f-Fermi} f(k^+,\vec
k_\perp)=\left(\exp\left[\frac{1}{k_BT}\left(\frac{k_{\rm on}^-+k^+}{2}
      -\mu\right)\right]+1\right)^{-1}  
\end{equation}
is expressed in terms of light front form momenta that are defined by
$\vec k_\perp=(k_x,k_y)$ and $k^\pm=k_0\pm k_z$. This treatment leads
to the same definition of blocking factors in the limit $T\rightarrow
0$ as given in Ref.~\cite{miller}. Note that the on $k^-$-shell light
front energy $k_{\rm on}^-=(\vec k_\perp^2+m(\mu,T)^2)/k^+$ depends on
$\mu$ and $T$. The (familiar) mass dependence is shown in
Figure~\ref{fig:mass}.

\subsection{Two-body case}
The technical difficulties including angular momentum in relativistic
many-body systems are well known. For the time being we average over
the spin projections which means that the spin degrees of freedom are
washed out in the medium. This will be improved while the
investigation proceeds along the lines suggested in
Ref.~\cite{BKW:98}. Also we expect antiparticle degrees of freedom to
be of minor importance for a zero range interaction on the light
front~\cite{Frederico:1992}.  The solution for the two-body propagator
$\tau(M_2)$ for a zero-range interaction is given
by~\cite{Frederico:1992}
\begin{equation}
\tau(M_2)=\left(i\lambda^{-1} - B(M_2)\right)^{-1}.
\label{eqn:prop2}
\end{equation}
where the expression for $B(M_2)$ in the rest system of the two-body system
is 
\begin{equation}
B(M_2)=-\frac{i}{(2\pi)^3} \int \frac{dx d^2k_\perp}{x(1-x)}
\frac{1-f(x,\vec k^2_{\perp})-f(1-x,\vec k_\perp^2)}{M_2^2-M_{20}^2} ,
\end{equation}
where $M_{20}^2=(\vec k_\perp^2+m^2)/x(1-x)$ and $x=k^+/P^+_2$.

\subsection{Three-body case}
The solution for the two-body propagator $\tau(M_2)$ is the input for
the relativistic three-body equation. The inclusion of finite temperature and
chemical potential is determined by the Dyson equations as explained,
e.g. in Ref.~\cite{Beyer:2001iz}. With the introduction of the vertex
function $\Gamma$ the equation becomes
\begin{eqnarray}
\lefteqn{\Gamma(y,\vec q_\perp) = \frac{i}{(2\pi)^3}\ \tau(M_2)
\int_{M^2/M_3^2}^{1-y} \frac{dx}{x(1-y-x)}}\nonumber\\
&&\int^{k_\perp^{\mathrm{max}}} d^2k_\perp
\frac{1-f(x,\vec k^2_\perp)
-f(1-x-y,(\vec k + \vec q)^2_\perp)}
{M^2_3 -M_{03}^2}\;\Gamma(x,\vec k_\perp) ,
\label{eqn:AGS}
\end{eqnarray}
where $m=m(\mu,T)$,
\begin{equation}
k_\perp^{\mathrm{max}}=\sqrt{(1-x)(xM_3^2-m^2)},
\end{equation}
and the mass of the virtual three-particle state in the rest system is
given by
\begin{equation}
M_{03}^2=\frac{\vec k^2_\perp+m^2}{x}
+\frac{\vec q^2_\perp+m^2}{y}
+\frac{(\vec k+\vec q)^2_\perp+m^2}{1-x-y} .
\end{equation}
The blocking factors, $1-f-f$ that appear in (\ref{eqn:AGS}) can be
rewritten as $\bar f\bar f -ff$, where $\bar f=1-f$ to exhibit the
particle and the hole blocking.  For $T\rightarrow 0$, $\bar f
\rightarrow\theta(k-k_F)$ that cuts the integrals below the Fermi
momentum.

\begin{figure}[t]
\begin{minipage}{0.48\textwidth}
\epsfig{figure=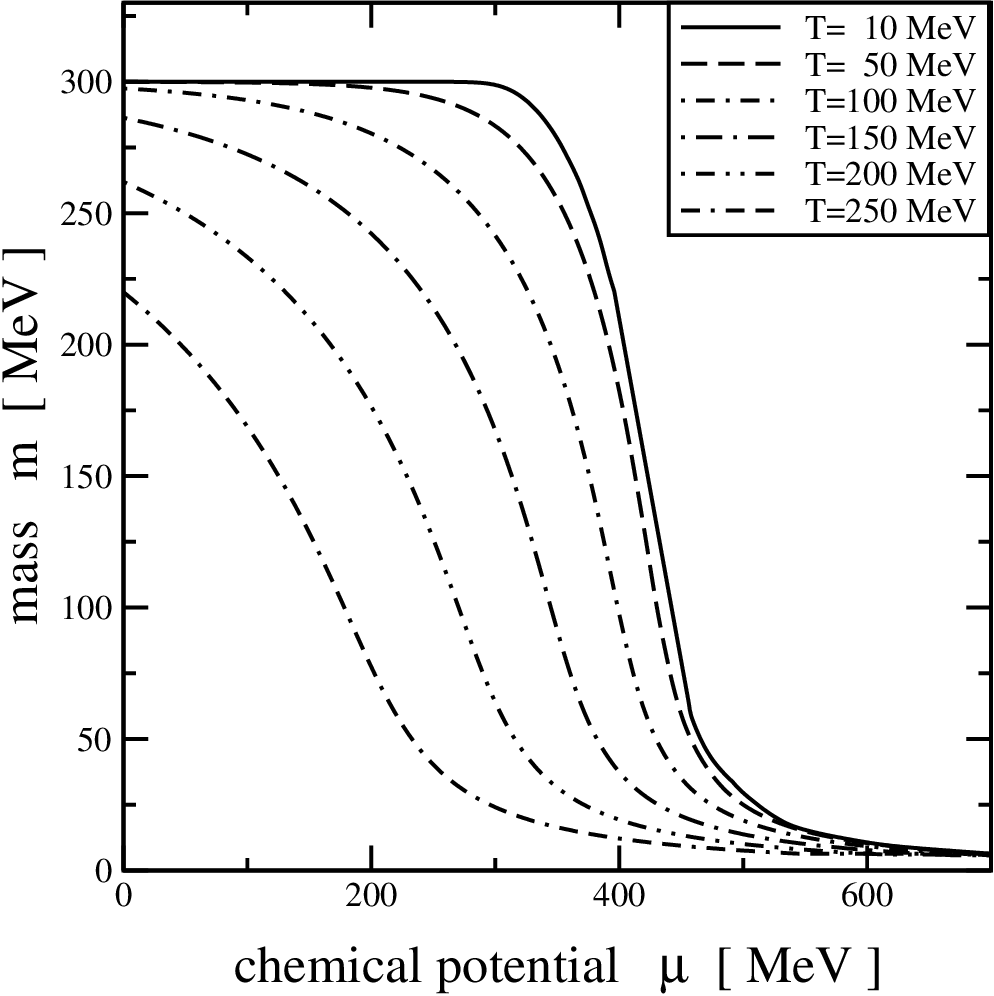,width=\textwidth}
\caption{\label{fig:mass} 
  Dependence of the mass on the (quark) chemical potential and
  temperature.~\vspace{5mm}}
\end{minipage}
\hfill
\begin{minipage}{0.48\textwidth}
\epsfig{figure=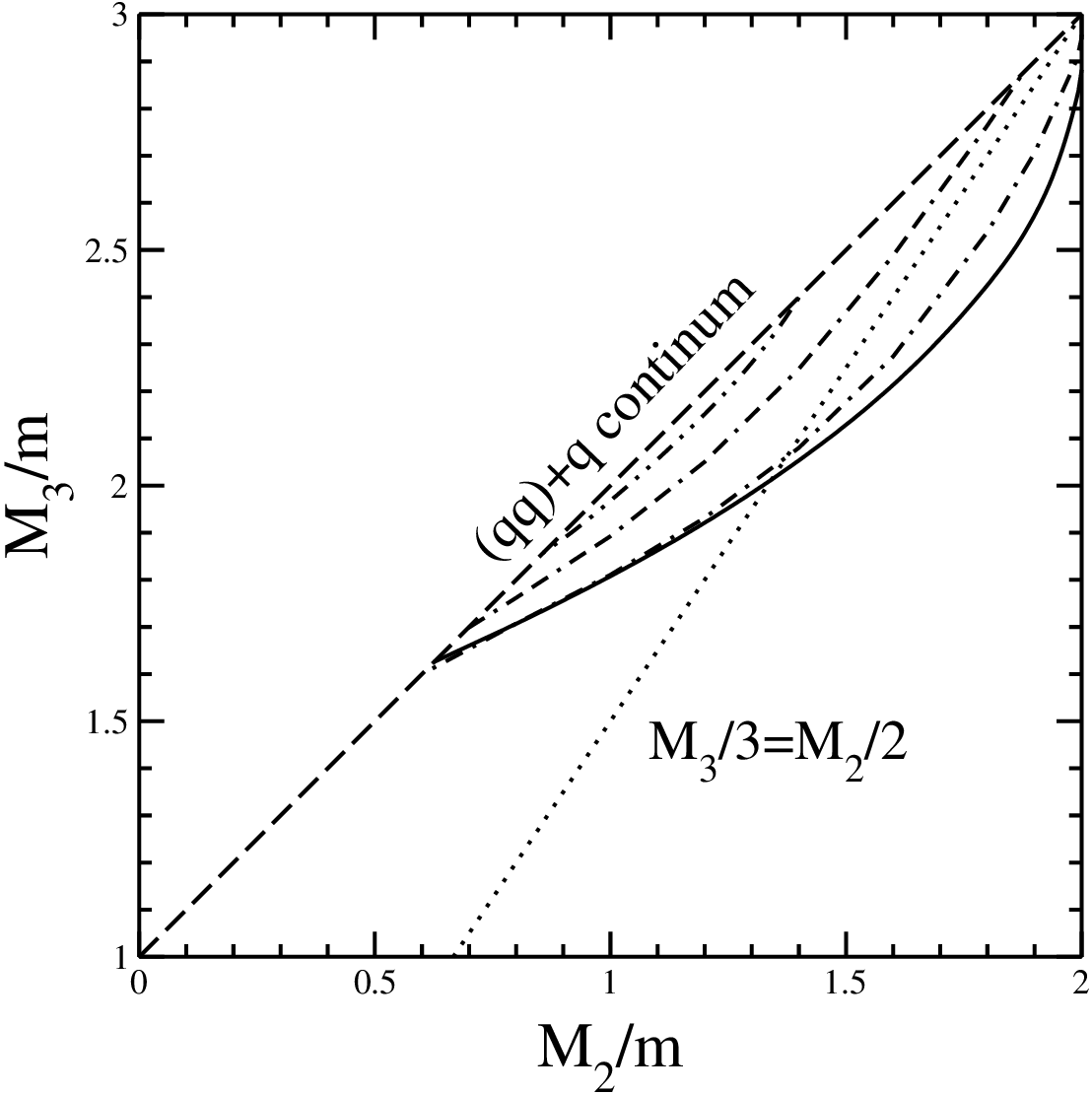,width=\textwidth}
\caption{\label{fig:M2M3} Masses of two-quark $M_2$ vs.
  three-quark $M_3$ bound states at $T=10$ MeV for
  different $\mu$. For further explanations see text.}
\end{minipage}
\end{figure}
\begin{figure}[t]
\begin{minipage}{0.48\textwidth}
\epsfig{figure=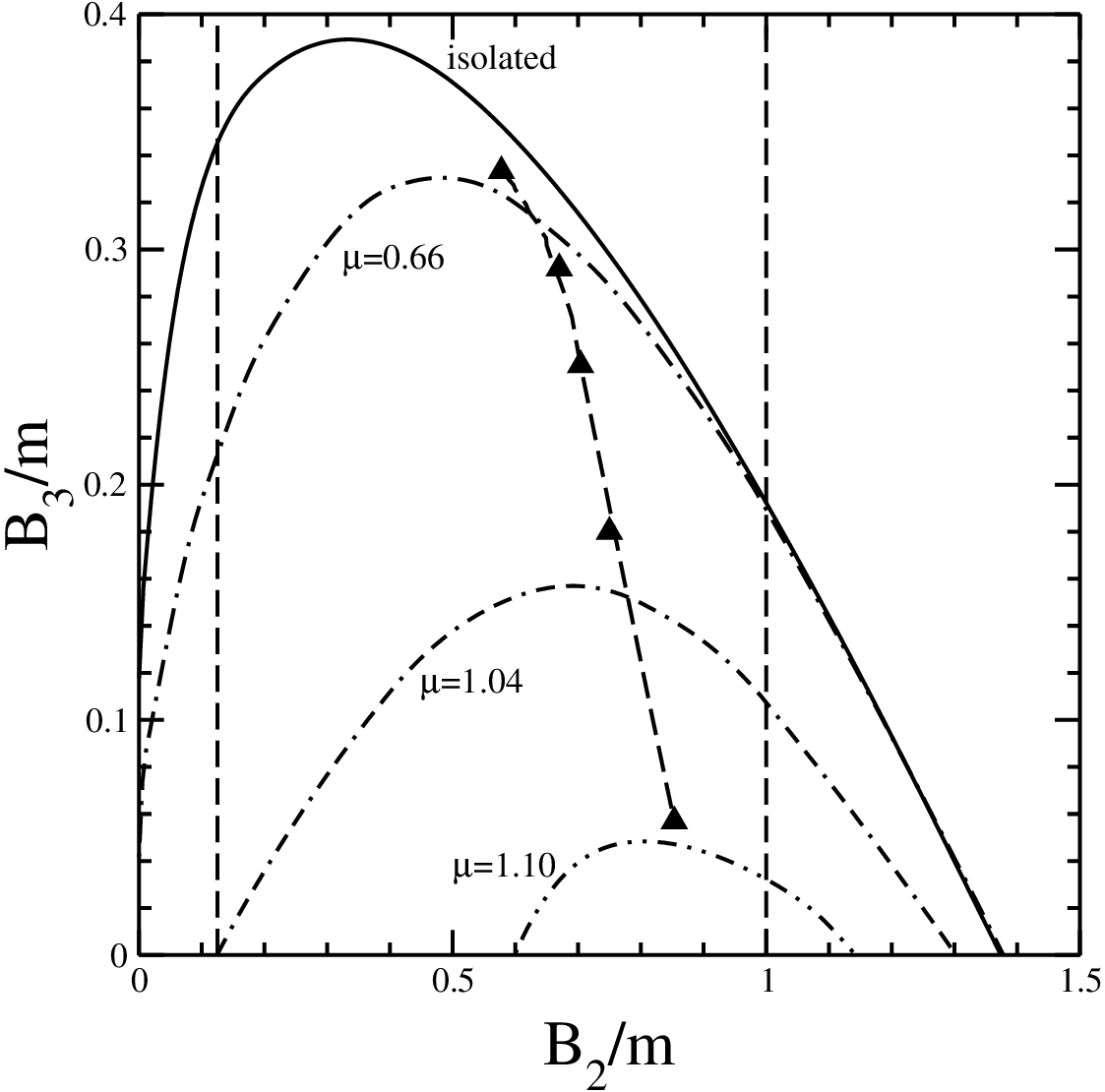,width=\textwidth}
\caption{\label{fig:medT10} Correlations between two-quark and
  three-quark binding energies in units of quark mass at $T=10$ MeV
  for different chemical potentials $\mu$, as indicated. Dashed with
  triangles \cite{Pepin}, further explanations see text.}
\end{minipage}
\hfill
\begin{minipage}{0.48\textwidth}
\epsfig{figure=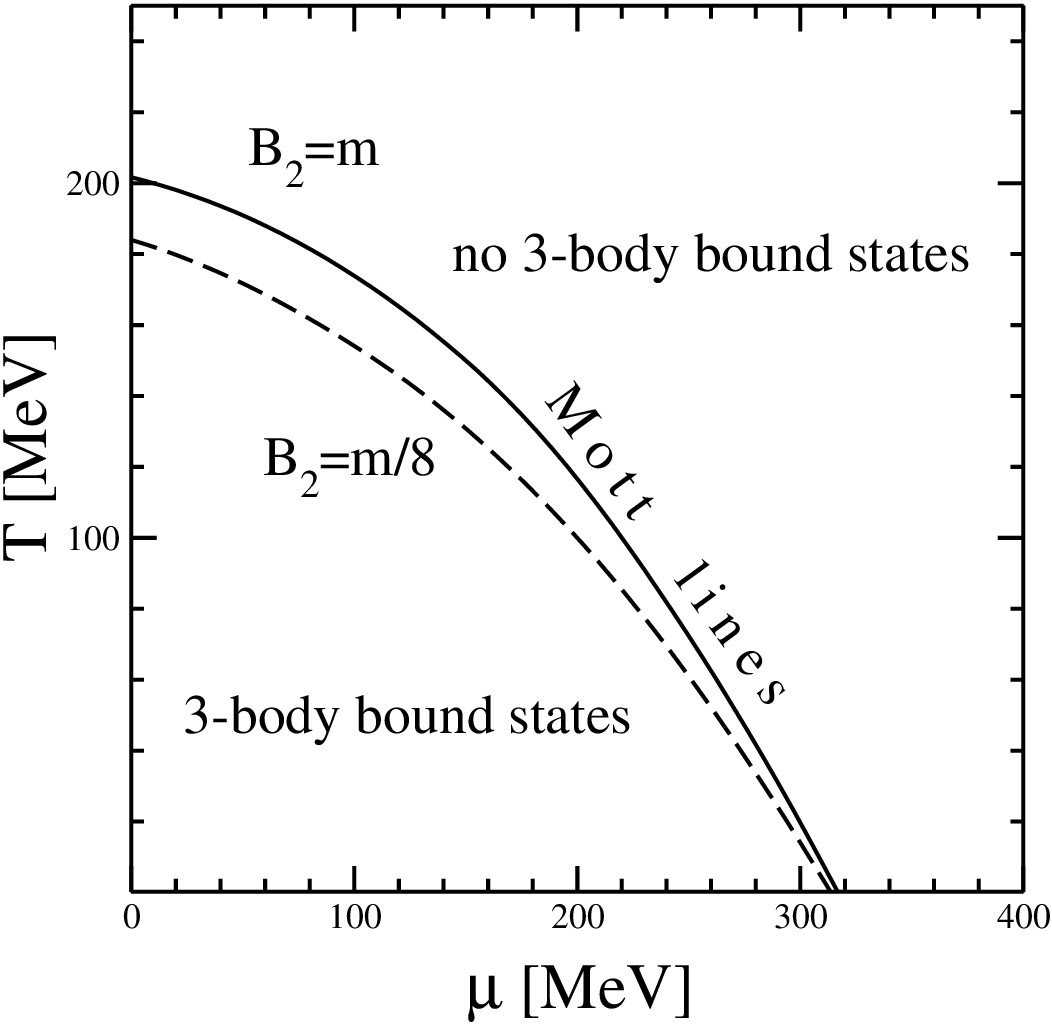,width=\textwidth}
\caption{\label{fig:mott}Mott lines for the three-body system at rest
  in the medium. For values of $T$ and $\mu$ below the Mott lines
  three-body bound states can be formed. Solid line for $M_2=m$,
  dashed line for $M_2=15m/8$.}
\end{minipage}
\end{figure}

\section{Results}

For the time being we assume a bound state $M_{2B}$ in the two-body
subsystem.  This can be relaxed while the investigation proceeds and
more realistic models are implemented. Our main focus is on effects
the medium has on the competition between two and three-quark states.
This is particularly relevant in the vicinity of the critical
temperatures. A first step towards this aim here is the investigation
of the Mott transition from the three-body bound state to the two body
(2+1) channel that differs from the three-nucleon
case~\cite{Beyer:1999zx}.  To this end, we vary $M_{2B}$ implicitly
choosing a particular model strength $\lambda$ in (\ref{eqn:prop2})
for each value of $M_{2B}$.  This is shown in Figure~\ref{fig:M2M3}.
For a given $M_{2B}$ the solid line reflects the corresponding
three-body bound state for the isolated system. For a temperature of
$T=10$~MeV the various dashed-dotted lines correspond to increasing
chemical potential (see Figure~\ref{fig:medT10} for the corresponding
values of $\mu$). The long dashed line is the two-body threshold.  In
a simple chemical picture, the equilibrium composition of the system
is dominated by three-body bound states below the dotted line and by
two-body states above this line (law of mass action).  In
Figure~\ref{fig:medT10} we show the binding energies
\begin{eqnarray*}
  B_3(\mu,T)&=&m(\mu,T)+M_{2B}(\mu,T)-M_{3B}(\mu,T)\\[-0.4ex]
  B_2(\mu,T)&=&2m(\mu,T)-M_{2B}(\mu,T).
\end{eqnarray*}
The solid line refers to the isolated case and the various
dashed-dotted lines again represent the in-medium results for
different chemical potentials at $T=10$ MeV.  In addition, we refer to
the NJL results (dashed line with triangles) given in
Ref.~\cite{Pepin} for $m=450$ MeV and $T=0$: The triangles are for
values of $\mu(T=0)\simeq 1.0,1.05,1.08,1.12,1.22$ (top to bottom) in
units of the respective $m(\mu,0)$.  The dashed vertical lines reflect
specific values for $M_{2B}=m$ and $M_{2B}\simeq 1.88m$ in our model
covering a wide range of $M_{2B}$ and for $T=10$ MeV. For a given
two-body binding energy and increasing chemical potential we find
weaker three-body binding.  This leads to the disappearance of the
three-quark bound state for a certain value of the chemical potential
which is known as Mott transition, $B_3(\mu_{\rm Mott},T_{\rm
  Mott})=0$. The values of $T$ and $\mu$ for which this transition
occurs is plotted in Fig.~\ref{fig:mott} for the two different models
given above. Clearly the behavior qualitatively reflects the
confinement deconfinement phase transition.

In conclusion, we have given a consistent equation for a relativistic
three-quark system in a medium of finite temperature and density. This
equation includes the dominant effects of the medium, viz. Pauli
blocking and self energy corrections. As many details need to be
improved, this first calculation shows that for a large range of
models the Mott transition agrees qualitatively with the phase
transition expected from other sources. Based on this approach it is
now possible to systematically investigate the influence of
three-quark correlations on the critical temperature and the onset of
color superconductivity at high density.

{\em Work supported by Deutsche Forschungsgemeinschaft.}

\end{document}